\documentclass{elsart} 
\usepackage{epsfig} 
\usepackage{bm} 
\usepackage{color}

\newcommand{\Onecol} 
{\begin{widetext} 
\onecolumngrid} 
\newcommand{\Twocol} 
{\end{widetext} \twocolumngrid}
 
\newcommand{\be}{\begin{equation}} 
\newcommand{\ba}{\begin{array}} 
\newcommand{\bea}{\begin{eqnarray}} 
\newcommand{\bfi}{\begin{figure}} 
\newcommand{\ee}{\end{equation}} 
\newcommand{\ea}{\end{array}} 
\newcommand{\eea}{\end{eqnarray}} 
\newcommand{\efi}{\end{figure}}

\begin{document}
\begin{frontmatter}
\title{Bursting in a Subcritical Hopf Oscillator with a Nonlinear Feedback}  
\author{Gautam C Sethia$^{1}$ and Abhijit Sen$^{2}$}
%\email{gautam@ipr.res.in}
%\author{Abhijit Sen}
%\email{abhijit@ipr.res.in}
\address{Institute for Plasma Research, Bhat, Gandhinagar 382 428, India}
\thanks{E-mail: gautam@ipr.res.in\\ $^2$ E-mail: abhijit@ipr.res.in}
\begin{abstract}
 Bursting is a periodic transition between a quiescent state and a state of repetitive spiking. The phenomenon is ubiquitous in a variety of neurophysical systems. We numerically study the dynamical properties of a normal form of subcritical Hopf oscillator (at the bifurcation point) subjected to a nonlinear feedback. This dynamical system shows an infinite-period or a saddle-node on a limit cycle (SNLC) bifurcation for certain strengths of the nonlinear feedback. When the feedback is time delayed, the bifurcation scenario changes and the limit cycle terminates through a homoclinic or a saddle separatrix loop (SSL) bifurcation. This system when close to the bifurcation point exhibits various types of bursting phenomenon when subjected to a slow  periodic external stimulus of an appropriate strength.   The time delay in the feedback enhances the spiking rate i.e. reduces the interspike interval in a burst and also increases the width or the duration of a burst. 
\end{abstract}
%\pacs{87.19.La, 05.40.Ca, 82.40.Bj, 87.18.Bb}  
\end{frontmatter}

\section{Introduction}
%\PARstart{T}{he} second National Conference on Nonlinear Systems 
When a system's activity alternates between a quiescent state and a state of repetitive spiking, the system is said to exhibit bursting behaviour \cite{izhikevich00,hoppensteadt,rinzel}. These large excursions exist because the system  is close to bifurcations from rest to oscillatory states. There are two bifurcations which characterize a burst; one which leads a rest state to repetitive spiking and the second which leads a spiking state to the rest state. As there are varieties of such bifurcations one finds a number of different types of bursts \cite{izhikevich00}. The bursting is ubiquitous in a variety of physical and biological systems, especially in  neural systems where it plays an important role in communication among neurons \cite{lisman97,izhikevich02,izhikevich03,kepecs04}. The spiking rate within a burst and also its duration are considered important for the reliability of neuronal communication.
\\
\indent
In this paper we describe a mathematical model which contains the necessary ingredients to not only produce different types of bursts but to also control the width and the spiking rate within a burst. This becomes possible due to introduction of a time delay in the self-feedback term of the system. We may point out here that the self-feedback can mock up a variety of physical effects. In a collection of coupled systems, the term can be regarded as a source term representing the collective feedback due to the rest of the systems. A similar single equation model as a paradigm for studying the collective dynamics (synchronization, death states etc.) of an ensemble of coupled oscillators has been proposed and discussed in depth in \cite{reddy00}. The time delays, arising from finite conduction/propagation speeds, can certainly be significant in situations where the systems communicate over some distance. The feedback is time delayed to account for such propagation delays of the signals.
\section{The Model}
 A topological normal form of the subcritical Hopf oscillator is given as:

\begin{equation}
\dot{z}(t)=(\mu+i(\omega+b\vert z(t)\vert ^{2})+\vert z(t)\vert ^{2}-\vert z(t)\vert ^{4})z(t) \label{original}
\end{equation} 
where $z=x+iy $ is a complex variable and $\mu$ is the bifurcation parameter. The frequency of the oscillations is determined by $\omega$ and $b\vert z\vert^{2}$. The parameter $b$ which is called the shear parameter, determines how the frequency depends on the amplitude of the oscillations. Canonical models derived from the above and similar normal forms have been widely used in the past for studying the bursting behaviour of neurons \cite{ermentrout86,ermentrout86a,izhikevich00a}. In what follows, we alter the above model to incorporate a nonlinear feedback term and then study its repercussions on the bifurcation dynamics and consequently on the excitability of the system.\\
\indent
The system  dynamics near Hopf bifurcation point is of intertest so for our present studies we assume the oscillator to be residing at the subcritical Hopf bifurcation point i.e. $\mu=0$. We add a nonlinear feedback term to the equation. The feedback is time delayed to incorporate the unavoidable time delays in signal transmission in any physical or biological system.  
In our model we have introduced a quadratic nonlinear feedback which is the simplest and the lowest order nonlinearity that provides an excitable behaviour to the model as we see later.
We denote the feedback strength by $k$ and the time delay by $\tau$. The modified oscillator equation can then be written as:
\begin{equation}
\dot{z}(t)=(i(\omega+b\vert z(t)\vert ^{2})+\vert z(t)\vert ^{2}-\vert z(t)\vert ^{4})z(t)-kz^{2}(t-\tau) \label{feedback}
\end{equation} 
We solve and carry out bifurcation analysis of this model equation rewritten in $x$ and $y$ coordinates.
The next section gives the different bifurcation scenario with the nonlinear feedback strength $k$ and the time delay $\tau$.\\
\indent
Most of the simulations in the present paper have been carried out using software package XPPAUT \cite{xppaut}  which also provides an interface to the continuation software package AUTO \cite{auto}. The bifurcation studies with time delay have been carried out using DDE-BIFTOOL \cite{dde}. 
%%%%%

\begin{figure}[tbh]
\centering
\includegraphics[width=0.7\columnwidth]{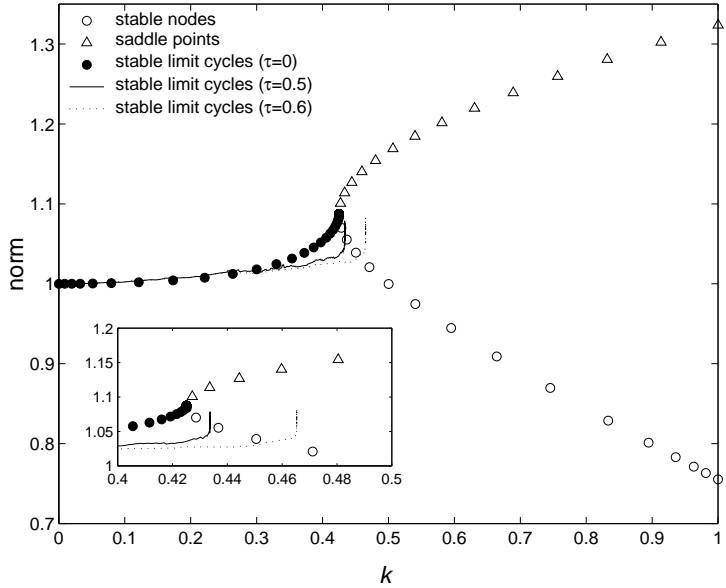}
\caption{Bifurcation diagrams as a function of the feedback strength $k$ with and without time delay $\tau$.}
\label{bif_k}
\end{figure}
\begin{figure}[tbh]
\centering
\includegraphics[width=0.7\columnwidth]{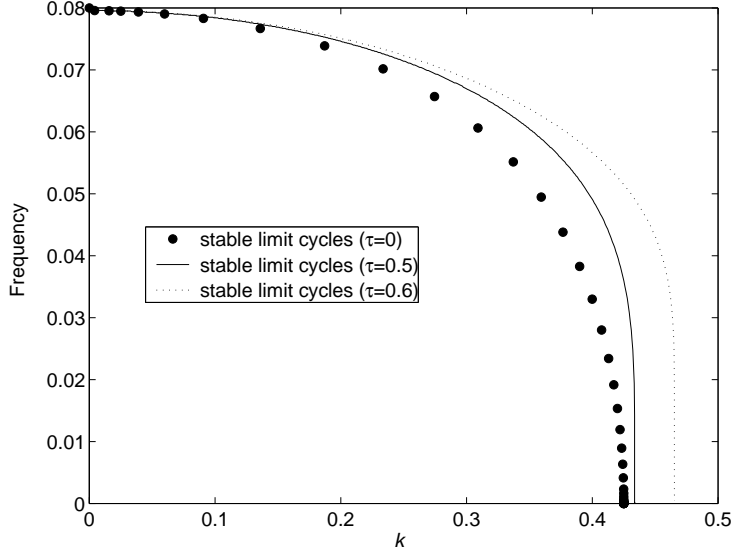}
\caption{The frequencies of the limit cycles tending to zero as the bifurcation is approached.}
\label{freq}
\end{figure}
\begin{figure}[tbh]
\centering
\includegraphics[width=0.7\columnwidth]{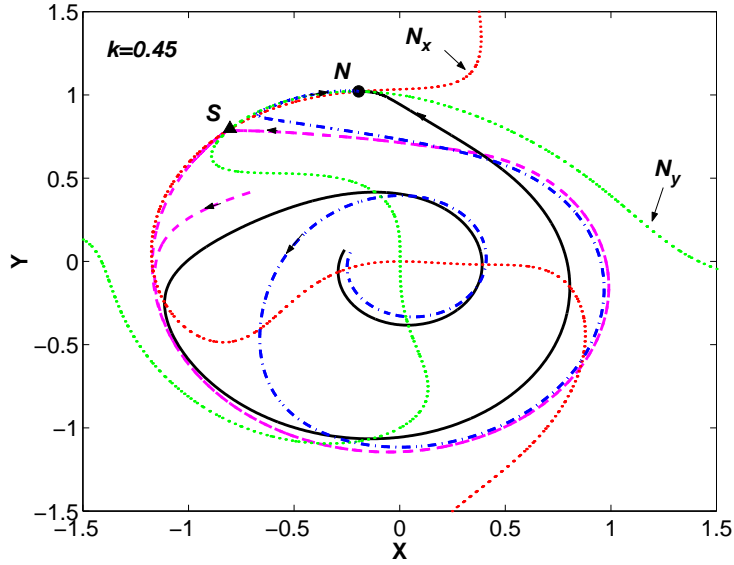}
\caption{Typical trajectories for various values of $\tau$ ($\tau =0:$ solid; $\tau=0.5:$ dash-dot; $\tau=0.57:$ dash) illustrating a saddle separatrix loop bifurcation beyond a threshold $\tau_t$. The nullclines (\textit{$N_{x}$,$N_{y}$}) along with the saddle (\textit{S} ) and the node (\textit{N} ) points for $\tau=0$ are also marked.}
\label{traj}
\end{figure}

\section{The Bifurcations}
\begin{figure}[tbh]
\centering
\includegraphics[width=0.7\columnwidth]{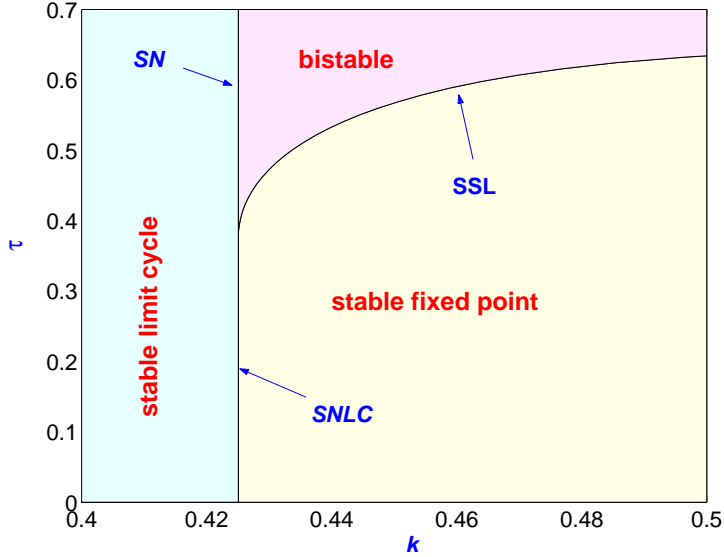}
\caption{Stability diagram in the parameter space of $k$ and $\tau$. Note the various bifurcation boundaries demarcating the different stability regions.}
\label{bif_ktau}
\end{figure}

We first carry out a detailed bifurcation analysis for our model system in terms of the feedback strength parameter $k$ for different values of the time delay parameter $\tau$.
Our results are shown in Fig. \ref{bif_k} where plots of the L2-norm ($\vert z\vert$) $vs$ $k$ are shown for $b=-0.5$ and $\tau=$ $0.0$, $0.5$ and $0.6$ respectively. The corresponding frequencies of the limit cycles are shown in Fig. \ref{freq} and one clearly notices frequencies tending to zero as the bifurcation point is approached. Going back to Fig. \ref{bif_k}, we find that in the absence of time delay ($\tau=0$) and for $k>0.42506$ there exist two fixed points - a stable node and a saddle. At $k=k_{c}=0.42506$ we have an SNLC bifurcation such that for $k<k_c$ one only has a stable limit cycle branch whose period tends to infinity at the bifurcation point (see Fig. \ref{freq}). The location of $k_c$ depends on the value of the shear parameter $b$ but the nature of the bifurcation diagram remains the same as long as $-1<b \leq -0.5$.  The presence of finite time delay in the feedback introduces profound changes in the bifurcation diagram of the system. As seen from Fig. \ref{bif_k} the stable limit cycle branch now extends beyond the SNLC bifurcation point of $k=k_c$ and gives rise to a bistable region over this extension. The point $k=k_c$ now marks a saddle node (SN) bifurcation point. The range of extension increases as a function of $\tau$. 
The existence of the limit cycle in the region beyond $k=k_c$ has a threshold character in $\tau$ i.e. for a given $k$ beyond $k_c$ the limit cycle appears only when $\tau$ exceeds a threshold value ($\tau_t=0.3759$). Some typical trajectories for different values of $\tau$ and a fixed value of $k>k_c$ are shown in Fig. \ref{traj}. For $\tau$ values below a threshold all trajectories end up on the stable node and there is no oscillatory behavior. At a certain critical value of $\tau$ one notices the emergence of a separatrix loop trajectory and the consequent onset of periodic behavior.
The dynamical origin of this behavior can be traced to the onset of an SSL bifurcation \cite{izhikevich00} occurring as the parameter $\tau$ is varied. In both the cases of bifurcation (SNLC and SSL), the period of the limit cycle becomes infinite or the frequency becomes zero at the bifurcation point as we have already seen in Fig. \ref{freq} for the cases shown in Fig. \ref{bif_k}. These bifurcations with their scaling laws have been beautifully explained by Strogatz \cite{strogatz}.\\
\indent
The two parameter bifurcation diagram in $k$ and $\tau$ space is shown in Fig. \ref{bif_ktau}. Different bifurcation branches as well as the regions of stable limit cycle/fixed point/bistable solutions are clearly demarcated. A system close to the meeting point of these three regions can exhibit four diffrent type of bursting \cite{izhikevich00}. A system can switch over from a steady state to the oscillatory state either  through a saddle-node on a limit cycle (SNLC)  or a through  saddle-node (SN) bifurcation whereas the system can switch back from the oscillatory to the steady state through SNLC or a saddle separatrix loop (SSL) bifurcation. In principle, this gives rise to the possibility of four different types of bursts to arise: SNLC/SNLC (Type II or parabolic bursting), SNLC/SSL (no name yet), SN/SNLC (triangular bursting), and SN/SSL (Type I or square-wave bursting); the first bifurcation refers to the steady state to the oscillatory state and the second one refers to the oscillatory to the steady state bifurcation. The parabolic (SNLC/SNLC) and the square-wave (SN/SSL)  bursting are quite common in neuronal modelling. We now check the feasibility of bursting behaviour of this model through a simulation work as described in the next section.

%%%%%%%%%%%%%%
\section{Simulation of bursting}
For the purpose of simulating bursts, we add a slow stimulus to the model (Equation \ref{feedback}) which can drive the system through bifurcations of equilibria and limit cycles. The stimulus chosen is $s(t)=\varepsilon e^{i\Omega t}$ where $\varepsilon$ is the amplitude and $\Omega$ is the frequency of a periodic signal. The amplitude and the frequency of $s(t)$ determines whether the system evokes no-spikes (subthreshold signal), single spikes, or bursts. We choose $\varepsilon=0.02$ and $\Omega=0.01$ such that the system gets across the bifurcation regions for all the four different feedback strengths $k$ chosen for the present simulation. The other parameter values are $b=-0.5$ and $\omega=1$. The time delay $\tau=0,0.3$ and $0.5$ for Figs. \ref{burst_1},\ref{burst_3}, and \ref{burst_2} respectiveley. The four feedback strengths ($k's$) chosen are such that $k=0.41$ and $0.42$ are in the oscillatory region and $k=0.43$ and $0.44$ are in the fixed point or the rest region. The stimulus added  plays the role of driving the system across the bifurcation boundaries.
\subsection{Bursting with no delay}
\begin{figure}[tbh]
\centering
\includegraphics[width=0.7\columnwidth]{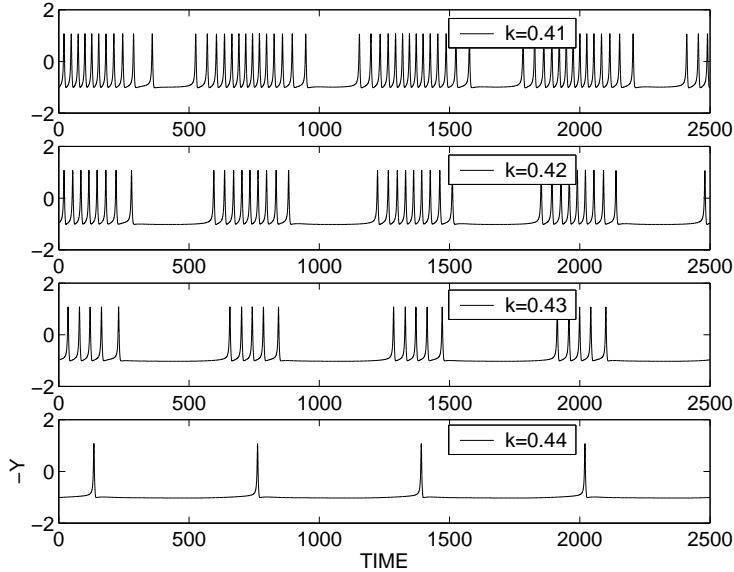}
\caption{"SNLC/SNLC" (parabolic) bursting in the absence of delay ($\tau=0.0$) for four different values of the feedback strenght $k$. Other parameters are $b=-0.5,\omega=1, \epsilon=0.02, \Omega=0.01$.}
\label{burst_1}
\end{figure}

We first simulate the case with zero delay ($\tau=0$) and the results are shown in Fig. \ref{burst_1}. The bottom curve correspond to the case when $k=0.44$ which is quite far from the SNLC boundary or the critical $k_c$ and the stimulus just manages to generate a single spike burst. As $k$ is decreased to a value of $0.43$ we get the parabolic bursts. Since the onset as well as the termination of the burst is due to SNLC, the period gets larger on both ends of the burst. The larger spacings on both ends of the burst is an indicator of that. The top two curves for $k=0.41$ and $0.42$ correspond to the case when the system is already in the oscillatory region and the stimulus pushes the system back and forth into the fixed point region to generate the bursts. The two cases, the one in which the system is in the fixed point region and the other in which the system is in oscillatory region to start with, are distinguishable by the fact that the system has shorter duration bursts in the first case and larger duration bursts in the second. In the first case the bursting activity correspond to a continuous state of quiescence or fixed point interruptedd by a spike or a cluster of spikes whereas in the second case it correspond to a continuous nonuniformly spaced spike train interrupted by short spells of quiescencee.
\subsection{Bursting with finite delay}
\begin{figure}[tbh]
\centering
\includegraphics[width=0.7\columnwidth]{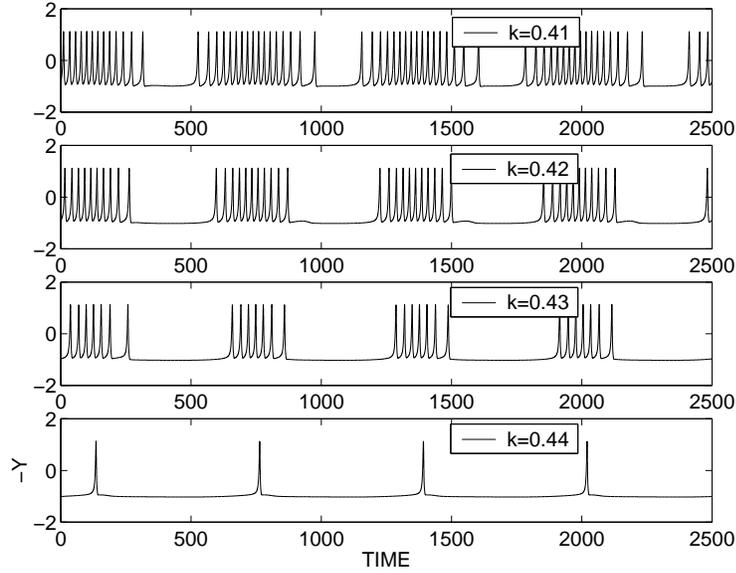}
\caption{"SNLC/SNLC" (parabolic) bursting with finite time delay ($\tau=0.3$) but less than the threshold value $\tau_t$ for the bistable region to appear. Other parameters are as in Fig.\ref{burst_1}}
\label{burst_3}
\end{figure}

\begin{figure}[tbh]
\centering
\includegraphics[width=0.7\columnwidth]{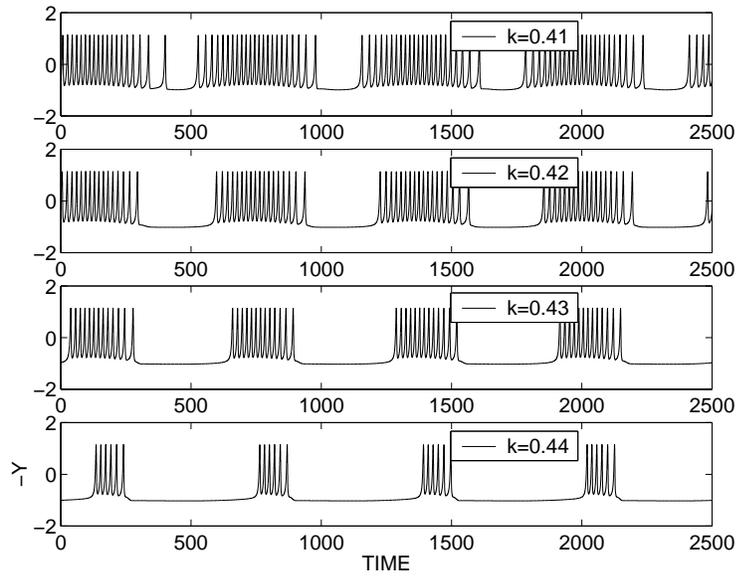}
\caption{"SN/SSL" (square-wave) bursting in the presence of a large enough time delay ($\tau=0.5$) to be in the bistable regime. Other parameters are as in Fig.\ref{burst_1}}
\label{burst_2}
\end{figure}

We now delay the feedback so as to be still in the monostable region (i.e. $\tau<\tau_t$),  and keep other parameters the same as in Fig. \ref{burst_1}, the results are depicted in Fig. \ref{burst_3}. One clearly notices an enhanced spike rate and an increase in the number of spikes per burst. These features get further enahanced if we increase the delay to get into the bistable region. Fig. \ref{burst_2} is for $\tau=0.5$ with other parameters the same as in Fig. \ref{burst_1}. The onset of the spiking is on the saddle-node off the limit cycle and hence spikes with a finite frequecies not tending to zero (see Fig. \ref{freq}) whereas the termination of the oscillatory state is through an SSL or a homoclinic bifurcation which slows down the spike rate on this end of the burst. One does notices these features in the simulated bursts. An increase or a decrease in the time delay within the bistable region too correspondingly  enhances or reduces the spike rate within a burst and its duration.
\section{Conclusions}
We have proposed a mathematical model for the spiking-bursting activity of a neuron. Eventhough the model is not based on physiology, it does simulate a number of features observed in neuronal bursting. When we examine the bifurcation scenario in two parameter space of the feedback strength and the time delay, we find the relevant bifurcations responsible for different types of bursting activities. The dynamics of our system is capable of generating highly nonlinear stable limit cycles, which mimic the spiking/bursting activity of a neuron, and a stable fixed point, which corresponds to a rest state. The system, more interestingly, also has a parameter regime where both these stable regimes (i.e. the oscillatory and the rest) coexist. Existence of such bistable regime is the reason for generation of square-wave bursting through hysteresis. The monostable rest and the oscillatory states, separated by a saddle-node bifurcation on a limit cycle, enables the generation of the parabolic bursts. The spike rate within a burst and its duration (or the width) gets enhanced by a time delay which is below threshold so as not to push the system in the bistable region. A large enough time delay pushes the system into the bistable region and we obtain square-wave bursts with  enhanced spike rate as well as the width of the square-wave bursts. \\

{\bf{Acknowledgments:}}
We are grateful to Prof. Bard Ermentrout for his useful tips on the use of XPPAUT and to Prof. Tatyana Luzyanina for her helpful suggestions on the use of DDE-BIFTOOL.
%\nocite{*}
\bibliographystyle{ncnsd}
%\bibliography{bibliography-file}

%% The bibliography file after composition (.bbl file) can be included
%% in the latex file as follows:

\end{document}